\documentclass[11pt,a4paper]{article}
\usepackage[ansinew]{luainputenc} 
\usepackage[hmargin=2cm, vmargin=2cm]{geometry}
\usepackage [T1]{fontenc} 
\usepackage{textcomp}
\usepackage{dsfont} 
\usepackage{mathrsfs} 
\usepackage{csquotes} 
\usepackage{amsmath,amsfonts,latexsym, amssymb,amsthm}
\usepackage{fullpage,url,booktabs}  
\usepackage[usenames, dvipsnames]{xcolor} 
\usepackage[colorlinks,citecolor=OliveGreen]{hyperref} 
\usepackage{array} 
\usepackage[sort&compress]{natbib}
\usepackage{stmaryrd} 
\usepackage{xspace}
\usepackage{tikz} 
\usepackage{stackrel} 
\usepackage{cancel} 
\usepackage{comment} 
\newtheorem* {proposition*}{Proposition}
\newtheorem* {theorem*}{Theorem}
\newtheorem* {corollary*}{Corollary}
\newtheorem* {definition*}{Definition}
\newtheorem* {lemma*}{Lemma}
\newtheorem* {assumption*}{Assumption}
\newtheorem* {remark*}{Remark}
\newtheorem {proposition}{Proposition}

\newtheorem {definition}{Definition}

\newtheorem {remark}{Remark}

\bibpunct{[}{]}{,}{n}{}{;} 

\newcommand{\Prob}{{\mathbb{P}}}

\newcommand{\ident}{\stackrel{\mbox{{\rm \tiny (def)}}}{=}}
\newcolumntype{C}{>{$}c<{$}}
\newcolumntype{L}{>{$}l<{$}}
\newcolumntype{R}{>{$}r<{$}}

\title{\bf
Solving satisfiability by Bayesian inference
}
\author {Michel Feldmann%
\thanks{Electronic address: michel.feldmann@polytechnique.org}
}
\date {}

\begin{document}

\maketitle
\abstract{
We demonstrate that any logical problem can be solved by Bayesian inference. In this approach, the distinction between complexity classes vanishes. The method is illustrated by solving the 3-SAT problem in polynomial time.
Beyond this, Bayesian inference could be the background of artificial neural network theory.
} 


\section{Introduction}
\label{introduction}
Statistical estimation is a technique usually employed to evaluate unknown parameters based on experimental data~\cite{wald, boyd}. The method consists in processing the data as a prior knowledge in the framework of Bayesian inference theory~\cite{jaynes} and then infer the most likely value of the unknown parameters. 
Now we propose to use the same technique to solve the satisfiability problem of a set of Boolean formulas, compelled by hypothesis to satisfy deterministic truth values. To this end, we regard the hypothesis as a prior knowledge and then infer the most probable assignment to the logical variables compatible with the prior, which constitutes in fact the solution of the  satisfiability problem. While using probability, this is in no way a randomized algorithm: Simply, we takes advantage of the tools of Bayesian inference theory to reformulate and eventually solve any discrete algorithm by a \emph{strictly deterministic procedure}.

As a matter of fact, based especially on the works by J. M. Keynes~\cite{keynes},  R.~T.~Cox~\cite{cox} and E.~T.~Jaynes~\cite{jaynes},  probability theory can be construed as an extension of the Aristotelian logic to cases where the variables are not wholly definite. Logical rules are thoroughly retained but they are expressed with continuous real-valued probabilities instead of discrete logical symbols and next processed by convex analysis~\cite{boyd}. Therefore, this technique can be an alternative to standard algorithms involving Boolean formulas, with the crucial advantage of being a powerful tool of optimization in the real-valued domain, which naturally bypasses the intractable combinatorial searches that plague discrete algorithms, making ineffective the standard algorithmic complexity classes.

The direct implementation of the method leads to construct a linear programming (LP) problem. It turns out that the dimension of the LP system can be polynomial in the size of the data set, irrespective of the complexity of any potential discrete algorithm. 
In other words, in this framework, the distinction between the complexity classes \textsf{P} and \textsf{NP} proves to be irrelevant. We will eventually address the so-called 3-SAT problem as emblematic of the \textsf{NP}-complexity class.

Using continuous variables instead of discrete symbols, we get the major result that \textsf{P}=\textsf{NP}. In conclusion, we emphasize that this is also the case of quantum computation and, what is more interesting, artificial neural networks.

\section{Background}
\label{background}
Algorithms are well defined procedures to derive an unknown output from a countable set of input data.

\subsection{Deterministic algorithms}

A precise definition of algorithms  was proposed by R. Karp~\cite{karp}. 
From the Church-Turing thesis, any such computational issues can be formulated in terms of language recognition problems accepted by some Turing machine. 
When the machine terminates after a finite number of steps the final state is either the accepting state or the rejecting state.
The theory of algorithmic complexity~\cite{goldreich}, due especially to  S. Cook~\cite{cook},  L.A.~Levin~\cite{levin} and   R.~Karp~\cite{karp} delimits different complexity classes, in particular the classes \textsf{P} and \textsf{NP}. 
Informally, class \textsf{P} is the ensemble of solving languages recognizable in polynomial time by a deterministic Turing machines, while class \textsf{NP} is the ensemble of problems that can be associated with a checking language of class \textsf{P}.

\subsection{Satisfiability problems}

We aim to solve the satisfiability problem on a finite ensemble of logical equations.  Therefore, there is at least one possible solving language, namely, the \emph{force brute} algorithm. Clearly, this procedure is guaranteed to terminate, i.e.,  the problem is decidable.
Consider a set of Boolean formulas defined on a finite set of discrete binary variables, so that each variable $\mathsf{X}_i:\{ 0,1\}\to\{ 0,1\}$ with  $i\in\llbracket 1,N\rrbracket$ can store just one bit of  information.
We adopt the gauge \enquote{$1$} for \enquote{valid} or \enquote{TRUE} and thus \enquote{$0$} for \enquote{invalid} or \enquote{FALSE}.   
For the sake of generality, we accept at the outset  a  number $N$ of variables greater than the number $N_\mathrm{in}$ of input variables so that $N=N_\mathrm{in}+N_\mathrm{aux}$, meaning that $N_\mathrm{aux}$ auxiliary variables are allowed to store intermediate results, like e.g. carry bits in addition. However, we will only solve completely the case where $N_\mathrm{aux}=0$, corresponding to the standard problem of \emph{satisfiability}. For clarity we will call \enquote{general satisfiability problem} the  case  $N_\mathrm{aux}>0$ and \enquote{strict satisfiability problem} the case $N_\mathrm{aux}=0$.

\subsection{Assignments, requirements, layers, states}
 We note   $\overline{\mathsf{X}}_i$ the negation $\neg\mathsf{X}_i$ of a variable $\mathsf{X}_i$,  and call \emph{literal} $\mathsf{Y}_i\in\{\mathsf{X}_i,\overline{\mathsf{X}}_i  \}$ a variable or its negation. A logical formula (or \emph{Boolean function}) is a mapping from $\{ 0,1\}^N\to\{ 0,1\}$.  Let $\mathcal{T}$ denote the set of Boolean functions on $\{ 0,1\}^N$. Given two logical formulas  $\mathsf{f}_1$ and $\mathsf{f}_2$, it is convenient to note  $(\mathsf{f}_1 ; \mathsf{f}_2)$ (with a semicolon) the conjunction $\mathsf{f}_1\wedge\mathsf{f}_2$ and  $(\mathsf{f}_1 , \mathsf{f}_2)$ (with a comma) the disjunction $\mathsf{f}_1\vee\mathsf{f}_2$. We name \emph{complete assignment}, $x\in\{ 0,1\}^N$, a full assignment of $0$ or $1$ to the $N$ variables and \emph{partial assignment} an assignment to less than $N$ variables.
For convenience, we will especially use the \emph{conjunctions} of literals to define the unknowns. 
Thus, we find helpful to name \emph{partial requirement} a conjunction of $\ell<N$ literals, for instance,  $({\mathsf{X}}_{i_1};\overline{\mathsf{X}}_{i_2};\dots;{\mathsf{X}}_{i_\ell})$ in $\{ 0,1\}^\ell$,
and   \emph{complete requirement} $\omega$, or \emph{state}, a conjunction of $N$ literals  in $\{ 0,1\}^N$, e.g., $\omega=({\mathsf{X}}_1;\overline{\mathsf{X}}_2;\dots;{\mathsf{X}}_N)$.
The set of partial requirements with exactly $\ell$ literals will be called the \emph{layer} of order $\ell$ or the $\ell$-layer. Therefore, there are $N$ distinct layers with $\ell>0$ and one additional layer with $\ell=0$ for the empty requirement.
A partial requirement is satisfiable by a partial assignment and a complete requirement $\omega$ by a complete assignment $x_\omega$, e.g., $(1; 0;\dots; 1)$. Clearly, there are $2^N$ different complete assignments and therefore $2^N$  complete requirements. Let $\Omega \ident \{ \omega\}$ denote the layer of order $N$, that is the set of complete requirements (or states). 

On the other hand, with up to $N$ variables, it is possible to construct card$(\mathcal{T})=2^{2^N}$ different Boolean functions, described, e.g., as  full disjunctive normal forms, i.e.,  disjunction of  complete requirements.  Thus, any Boolean function $\mathsf{f}$  is represented by a disjunction $(\omega_1,\omega_2,\dots,\omega_k)$ of $0\le k\le 2^N$ complete distinct requirements $\omega_i$. The tautology corresponds to $k=2^N$ and the empty function, $\mathsf{f}=\varnothing$, to $k=0$. 

General satisfiability and strict satisfiability problems can be formulated in the same basic form, irrespective of the status of the variables, either input variables or auxiliary variables. However, the crucial difference is that the truth table has only $2^{N_\mathrm{in}}<2^N$ independent entries in the first case. Of course, it would be possible in principle to formulate the problem with just the input variables but in general the size of the data set would increase exponentially.
In standard form, we aim to solve the following computational problem: 
\vspace{2mm}

\textsc{Problem} : Satisfiability of a set of $N=N_\mathrm{in}+N_\mathrm{aux}$ Boolean variables, composed of $N_\mathrm{in}$ input variables and  $N_\mathrm{aux}$ auxiliary variables.
The  case  $N_\mathrm{aux}>0$ is termed \textsc{general satisfiability problem} and the case $N_\mathrm{aux}=0$ is termed \textsc{strict satisfiability problem}. 

\textsc{Input}: A set of $n$ requirements with at most $\ell_\mathrm{max}$ layers over the set of $N$ variables,   a set of $m$ disjunctions of distinct requirements from the $n$ requirements. 

\textsc{Property}: The $m$ disjunctions of requirements are each compelled to satisfy a particular truth value. 

\vspace{2mm}

We will first pose the problem of general satisfiability irrespective of  $N_\mathrm{aux}$. Next we will completely solve the problem of strict satisfiability with $N_\mathrm{aux}=0$ and eventually detail this last issue specifically for the 3-SAT problem.

\subsection{ Kolmogorov probability space}
It turns out that the logical problem can be easily expressed in terms of probability.

Define as basic sample set the layer $\Omega = \{ \omega\}$ of all  $2^N$ states $\omega$. Since the cardinality of $\Omega$ is finite, the set of events is the power set $\mathfrak{P}(\Omega)$, of cardinality $2^{2^N}$, that is a sigma-algebra,  identical to  the ensemble of all Boolean functions $\mathcal{T}$. In other words, \emph{every Boolean function is an event of the sigma-algebra}. Next, we have to define a probability measure $\Prob$ on $\mathcal{T}$.
To this end, the logical hypotheses will be considered henceforth as a Bayesian \emph{prior}, say $({\Lambda})$.
Given that the prior $({\Lambda})$ has to be satisfied, \emph{the probability of any event will be conditional on $({\Lambda})$}.
Finally, the Kolmogorov probability space associated with the prior $(\Lambda)$ is $(\Omega, \mathcal{T}, \Prob)$. 

The actual probability distribution $\Prob$ is thus determined by the prior.
In general, there is a continuous set of distributions compatible with a prior $(\Lambda)$, while we search specifically for the deterministic solutions. 
Compared to a discrete algorithmic approach, the difficulty is now shifted to the determination of these solutions. 
We could tackle the problem by optimization but we will prove that no optimization is actually necessary for problems of strict satisfiability.

\subsection{Notation}
Throughout this paper, we will specifically name $\emph{unknowns}$ the conditional \emph{probability} of complete or partial requirements,  not to be confused with variables or Boolean functions subject to randomness.
Except when mentioned otherwise, we will use a shorthand to describe the unknowns, 
namely  
 $\Prob(i)$  for $\Prob({\mathsf{X}_i}=1|\Lambda)$,  
 $\Prob(-i)$ for $\Prob({\neg\mathsf{X}}_i=1|\Lambda)$, 
 $\Prob(i;-j)$ for $\Prob(\mathsf{X}_i\wedge{\neg\mathsf{X}}_j=1|\Lambda)$,
 $\Prob(i,-j)$ for $\Prob(\mathsf{X}_i\vee\neg\mathsf{X}_j=1|\Lambda)$,  
etc. (for $i,j\dots\in\llbracket 1, N\rrbracket$).
Similarly, we will use $\Prob(\omega)$ for $\Prob(\omega=1|\Lambda)$.
We will call \emph{partial probability} the probability of an unknown with less than $N$ literals, e.g. $\Prob(i;-j)$ and \emph{complete probability} the probability of an unknown $\Prob(\omega)$ with $N$ literals.  An unknown  labeled $k$ without  further detail will be denoted by $p_k$, e.g., we may have $p_k= \Prob(i;-j)$. An array of unknowns will be denoted by $p=(p_k)$.

\subsection{Universal equations.}
The rules of  logic reflected in the probability laws~\cite{cox}, can especially be expressed in the following universal relations:
\begin{align}
\label{Gconstraintsl}  \Prob( i_1; i_2;\dots;i_\ell) &\ge 0\\
\label{negationl} \Prob( i_1, i_2,\dots,i_\ell)&=1-\Prob( -i_1; -i_2;\dots;-i_\ell)\\
\label{simplet} \Prob(i) +\Prob( -i)\quad&=1\\
\label{tripletl} \Prob( i_1; i_2;\dots;i_\ell) &=\Prob( i_1; i_2;\dots;i_\ell;i_{\ell+1})+\Prob( i_1; i_2;\dots;i_\ell;i_\ell;-i_{\ell+1})
\end{align}
where $i, i_1,i_2,\dots i_{\ell+1}$ are signed integers and $|i_1|,| i_2|,\dots, |i_{\ell+1}|\in\llbracket 1,N \rrbracket$ are distinct.

Eq. (\ref{tripletl}) establishes a connection between the layer of order $\ell$ and the adjacent layer of order $\ell+1$.
 It is easy to establish that we have $\binom{N}{1}$ distinct equations like Eq. (\ref{simplet}),   $4\binom{N}{\ell}$ distinct equations like Eq. (\ref{tripletl}) with $\ell=2$ or  $12\binom{N}{\ell}$ equations with $\ell=3$, etc. 
Accounting for Eqs.~(\ref{simplet},  \ref{tripletl}), Eq.~(\ref{Gconstraintsl}) implies that
\begin{equation}
\label{bounds}
\Prob( i_1; i_2;\dots;i_\ell) \le 1.
\end{equation}
Also, from Eq. (\ref{tripletl}), the normalization equation of one literal Eq. (\ref{simplet}) can be formulated for a conjunction of $\ell$ distinct literals $\Prob( i_1; i_2;\dots; i_\ell)$ as a sum of $2^\ell$ terms as
\begin{equation}
\label{normel}
\sum_{2^\ell\mathrm{~terms}} \Prob(\pm i_1;\pm  i_2;\dots;\pm  i_\ell) = 1.
\end{equation}
A \emph{deterministic distribution} $\Prob$ is characterized by a state $\omega_0\in\Omega$ of the sample set so that
\begin{equation}
\Prob(\omega)=
\begin{cases}
1\quad\mathrm{~if~}\omega=\omega_0\\
0\quad\mathrm{~otherwise}
\end{cases}
\end{equation}
In this case, the probability of every Boolean function $\mathsf{f}$ in the sigma-algebra $\mathcal{T}$ is also deterministic, i.e., equal to 0 or 1. As a result, the probability distribution of any requirement is \emph{separable}, that is, a joint combination of independent distributions, as expressed by the following proposition:
\begin{proposition}
\label{propseparability}
In the deterministic realm, the probability $\Prob(i_1;i_2;\dots; i_\ell)$ of any requirement regarded as a joint distribution is always separable,  i.e.,
\begin{equation}
\label{separability}
\Prob(i_1;i_2;\dots; i_\ell)= \Prob(i_1)\cdot \Prob(i_2)\dots \Prob(i_\ell) 
\end{equation}
where $i_1,i_2,\dots i_\ell $ are signed integers and $|i_1|,| i_2|,\dots |i_\ell| \in\llbracket 1,N \rrbracket$ are distinct.
\end{proposition}
\emph{Proof.}  For a deterministic distribution, each partial probability is only 0 or 1. By defininition $\mathsf{X}_{-|i_k|}$ is the negation of $\mathsf{X}_{|i_k|}$. Now, by simple inspection, the two sides of Eq.~(\ref{separability}) are both equal to 1 if and only if all literals $\mathsf{X}_{i_k}$ are TRUE, otherwise they are both equal to 0. $\Box$

\section{Formulation of the Bayesian problem}
\label{LP}
Assume that we are given a formal description of a specific logical problem composed of a specific set of Boolean functions compelled to be either valid or invalid.
  It will be transcribed into a Bayesian prior, $(\Lambda)$, that is a set of linear equations subject to constraints. It is convenient to distinguish between \emph{specific} equations and \emph{consistency} equations.

\subsection{Specific equations}
A first set of equations, called \emph{specific equations}, is directly derived from the  specific Boolean hypotheses.
Technically, they are incorporated into the prior by assigning a probability of $1$  to events (or logical formulas) compelled to be valid and a probability $0$ to events compelled to be invalid. 
By simple inspection, any Boolean function is so naturally encoded as a \emph{linear} specific equation. For instance, a partial requirement $({\mathsf{X}}_i;\overline{\mathsf{X}}_j;{\mathsf{X}}_k)$, compelled to be valid or invalid in the Boolean algebra, is trivially encoded as $\Prob(i;-j;k)=1$ or $0$ respectively. A disjunction of disjoint expressions is encoded as a sum of probabilities.

It is convenient to consider the set of unknowns as composed only of \emph{partial probabilities}, i.e, probability of partial requirements, as opposed to a mixing of conjunctions and disjunctions.
A conjunction of expressions compelled to be valid may optionally be broken down into many distinct expressions compelled separately to be valid.
If necessary, we can switch a valid (resp. invalid) event to its negation, which is then compelled to be invalid (resp. valid). For instance, from Eq. (\ref{negationl}), the probability $\Prob(i,-j,k)=1$ of the valid disjunction $({\mathsf{X}}_i,\overline{\mathsf{X}}_j,{\mathsf{X}}_k)$ can be switched to the partial probability $\Prob(-i;j;-k)=0$ of the invalid partial requirement $(\overline{\mathsf{X}}_i;{\mathsf{X}}_j;\overline{\mathsf{X}}_k)$.   

\begin{definition}[Specific equations]
The specific equations are the set of linear equations reflecting directly the logical hypotheses in terms of partial probabilities.
\end{definition}

Starting from the logical hypothesis we simply encode each Boolean expression.
It is crucial to formulate the problem by using  a minimum set of unknowns, that is those explicitly displayed in the logical hypotheses.  Nevertheless,  to ensure consistency,  we need to add some extra unknowns, so that we will finally use a larger set of working unknowns. This is detailed in the following section (\ref{relevantp}).

\subsection{Working unknowns}
\label{relevantp}
Assume that the prior has been translated into an ensemble of specific equations involving a set of partial probabilities with a maximum of $\ell_\mathrm{max}$ literals per requirement, and thus a maximum of $\ell_\mathrm{max}$ layers.
We need to ensure that these unknowns depict genuine probabilities, i.e., that  the universal links between the layers hold. This  generally requires increasing the initial number of unknowns, leading to construct a set of \enquote{working unknowns}.

Indeed, each initial unknown,  for instance $\Prob(i)$, entails the logical consequence $\Prob(i) +\Prob(-i) =1$. We will name $\Prob(-i) $ a \emph{variant} of $\Prob(i) $. Similarly, the initial unknown $\Prob(i;j) $ entails the logical consequence  $\Prob(i) = \Prob(i;j)+\Prob(i;-j)$ and we will also name $\Prob(i) $ and $\Prob(i;-j)$ \emph{variants} of $\Prob(i;j)$. 
 It is convenient to call \enquote{positive unknown}, e.g. $\Prob(i;j;k)$ with  $i, j, k>0$ for $\ell=3$, an unknown composed of only \enquote{positive indexes}, $i, j, k$. 

\begin{definition}[Variant, positive variant]
The variants of an unknown are either the unknown itself or the unknowns obtained recursively by removing one literal of the initial unknown or by switching a literal into its negation. A positive variant is a variant composed of only direct literals and no negation, that is labeled with positive indexes.
\end{definition} 
 
 For instance, for the unknown $\Prob( i; j; k)$ there are $8$ variants in the 3-layer, namely, $\Prob(\pm i;\pm j; \pm k)$. More generally, for an unknown of the $\ell$-layer, there are $2^\ell$ variants among which one positive variant in the $\ell$-layer.  Next, there are $\binom{\ell}{\ell-1}=\ell$ positive variants in the $(\ell-1)$-layer, etc.

In practice, starting from an initial unknown of $\ell$ literals involved in a specific equation, we first derive the initial positive variant of $\ell$ literals and next the other positive variants of $\ell-1, \ell-2, \dots$ literals by removing successively all literals. Finally, we obtain all variants by switching any literal into its negation.
 For example, for each initial unknown of 3 literals $\Prob(i;j;k)$, the variants are  $\Prob(\pm i; \pm j ; \pm k)$, $\Prob(\pm i; \pm j)$, $\Prob(\pm j; \pm k)$, $\Prob(\pm k; \pm i)$, $\Prob(\pm i)$, $\Prob(\pm j)$, $\Prob(\pm k)$. i.e. $2^3 +\binom{3}{2} \times 2^2 +\binom{3}{1}\times 2^1= 3^3-1 =26$ variants.
 
More generally, for each working unknown of $\ell$ literals we have $3^\ell -1$ variants.
This number is then independent of the number $N$ of variables but exponential in the number $\ell_\mathrm{max}$ of layers used in the specific equations. From the list of partial probabilities involved in all specific equations, we can list all the variants. At last, we have naturally to remove the duplications. 

It is crucial that the maximum number of  layers $\ell_\mathrm{max}$  be independent of $N$. For instance, this maximum is $\ell_\mathrm{max}=3$ in the 3-SAT problem.   Under this condition, the total number of  working unknowns is polynomial in the size of the input data. 
 
 \begin{definition}[Working unknowns]
 The set of working unknowns is the ensemble of  partial probability variants involved in the specific equations.
 \end{definition}
 \begin{proposition}
 \label{numbunknowns}
 When the number of layers $\ell_\mathrm{max}$ involved in the specific equations is independent of $N$, the total number of  working unknowns is polynomial in the size of the input data.
 \end{proposition}

We will show that this set of working unknowns is sufficient to ensure the consistency of the formulation, as expressed by Proposition~(\ref{propdeterministic}) below.

\paragraph{Labeling the working unknowns.} When all the variants have been derived it is convenient to \emph{label the working unknowns in a single sequence} of (say) $n$ unknowns, $\{p_r\,|\,r\in\llbracket1,n\rrbracket\}$, where for example $p_r$ may stand for $\Prob(2;-3)$, meaning that $r$ is the label of $\Prob(2;-3)$ in this single sequence. 

\subsection{Consistency equations}
\label{compatibilityeq}
To ensure consistency, the logical link between the working unknowns and specifically between their variants must be made explicit.  
Precisely, for any unknown of the $(\ell+1)$-layer, the consistency equations express their universal links with the variants of the adjacent $\ell$-layer, Eqs. (\ref{simplet}, \ref{tripletl}).
They are easily derived from the list of all working positive unknowns.

 For instance,  for each  working positive unknown of 3 literals $\Prob(i;j;k)$, the consistency equations read 
\begin{align}
\label{triplet1}
\begin{aligned} 
\Prob(\pm i;\pm j)= \Prob(\pm i;\pm j ; k) +  \Prob(\pm i;\pm j ; -k) \\
\Prob(\pm j;\pm k)= \Prob(\pm j;\pm k; i) + \Prob(\pm j;\pm k ; -i) \\
\Prob(\pm k;\pm i)= \Prob(\pm k;\pm i ; j) + \Prob(\pm k;\pm i ; -j) 
\end{aligned}
\end{align} 
The number of equations is $\binom{3}{2}\times 2^{2}= 3\times 4=12 $.

More generally, for each  working positive unknown of $\ell$ literals we have $\binom{\ell}{\ell-1}\times 2^{\ell-1}= \ell\times 2^{\ell-1}$ consistency equations.
 \begin{definition}[Consistency equations]
 The {consistency equations} are the universal equations Eq. (\ref{simplet}, \ref{tripletl}) which link the adjacent layers.  
 \end{definition}
 For a fixed maximum number of layers, $\ell_\mathrm{max}$, the number of consistency equations is thus polynomial in the size of the input data. The total number of consistency equations is obtained after removing the possible duplications. 
 \begin{proposition}
 \label{numcomp}
  When the number of layers in the specific equations, $\ell_\mathrm{max}$, is independent of $N$, the number of consistency equations is polynomial in the size of the input data.
 \end{proposition}
 Again, this set of consistency equations is sufficient to ensure the consistency of the formulation, as expressed by Proposition~(\ref{propdeterministic}) below.

\section{Resolution of the satisfiability problem}
Collecting both the specific equations and the consistency equations, the prior is translated into a linear system. Let $n$ be the number of working unknowns an $m$ the total number of equations.

\subsection{Linear programming formulation}
We obtain a linear programming (LP) problem in stack variables~\cite{murty} defined in a convenient real-valued space $\mathbb{R}^n$  in the form,

\begin{align}
\label{lpeq}
\begin{aligned}
Ap&=b\\
\mathrm{subject~to~~~} p&\ge 0
\end{aligned}
\end{align}
where $p=(p_i)$ (with $i\in\llbracket 1, n\rrbracket)$ is a real unknown vector, $A=(\mathrm{a}_{j,i})$ (with $j\in\llbracket 1, m\rrbracket)$ a real matrix of $n$ columns and $m$ rows, and finally $b=(b_j)$  a real vector, while $p\ge 0$ stands for $\forall i, p_i\ge 0$. 
From Propositions (\ref{numbunknowns}, \ref{numcomp}), $n$ and $m$ are polynomial in the size of the input, i.e. in general, in the number of variables $N$.

Usually, for non trivial problems, the rank of the matrix $A$ is less than $n$ and thus, there is a continuous set of solutions.
This arises specifically when the problem accepts several solutions.
Now, we need to complete the computation by solving the LP problem. 
A feasible solution is a real-valued vector of unknowns, $p$,  that satisfies the prior $(\Lambda)$, that is Eq. (\ref{lpeq}), and therefore defines a probability distribution $\Prob$ on the set of working unknowns.

The only genuine solutions are of course deterministic. Thanks to the consistency equations, let us show that a deterministic solution on the set of working unknowns is also a deterministic solution on the full sigma-algebra $\mathcal{T}$. 

\begin{remark}
\label{rmq1}
The consistency equations specialize the LP system into a particular LP system which can be called \enquote{Bayesian}, in the sense that its solutions are probability arrays, that is to say whose entries are always in the interval $ [0 , 1] $.
\end{remark} 
Technically, it can be shown \cite{mf5} that the locus of the solutions in $\mathbb{R}^n$ is a compact simplex as opposed to a general polytope.

\subsection{Deterministic solutions}
From Proposition (\ref{propseparability}), the deterministic solutions on the full sigma-algebra $\mathcal{T}$ are separable.
It turns out that a deterministic solution restricted to the set of working unknowns is separable on  $\mathcal{T}$ as well.

\begin{proposition}
\label{lemmedeterministic}
A deterministic distribution over the set of working unknowns is separable.
\end{proposition}
\emph{Proof.}
Consider a feasible solution of Eq. (\ref{lpeq}) in which all working unknowns are deterministic, i.e., equal to $0$ or $1$.
Let us show that they are also separable, i.e., that Eq.~(\ref{separability}) holds specifically over the set of working unknowns. Taking into account the consistency equations, we proceed by induction on the number $\ell$ of literals per unknown,  for $1\le\ell\le\ell_\mathrm{max}$.

If $\ell=1$, the proposition is trivial.

Assume that Eq. (\ref{separability}) holds up to $\ell$ literals per unknown, i.e., $\Prob(i_1;\dots; i_\ell) = \Prob(i_1)\dots\Prob(i_\ell)$. 
Suppose that $\Prob(i_1;\dots; i_\ell;i_{\ell+1})$ is also a deterministic working unknown of $\ell+1$ literals. Then by consistency, $\Prob(i_1;\dots; i_\ell)$  is the sum of two non-negative terms, $\Prob(i_1;\dots; i_\ell;\pm i_{\ell+1})\ge 0$:  
\begin{equation}
\label{consisell}
\Prob(i_1;\dots; i_\ell)= \Prob(i_1;\dots; i_\ell;i_{\ell+1})+\Prob(i_1;i_2;\dots; i_\ell;-i_{\ell+1})
\end{equation}
From Eq. (\ref{consisell}), if $\Prob(i_1;\dots; i_\ell)=0$ then the two terms $\Prob(i_1;\dots; i_\ell;\pm i_{\ell+1})=0$ as well,  so that Eq. (\ref{separability}) holds for $\ell+1$ literals.
If $\Prob(i_1;\dots; i_\ell)=1$ then $\Prob(i_1)=1$, \dots,  $\Prob(i_\ell)=1$ which leaves two possibilities, either  $ \Prob(i_1;\dots; i_\ell; i_{\ell+1})=0$ and  $\Prob(i_1;\dots; i_\ell;- i_{\ell+1})=1$ or vice versa. Since $ i_{\ell+1}$ is a signed integer,  suppose for definiteness that the first possibility applies. A priori, we have still four cases 
\begin{center}
\begin{tabular}{CCCC}
\mathrm{case}&\Prob(i_1;\dots; i_\ell)&\Prob(i_{\ell+1})&\Prob(i_1;\dots; i_\ell;i_{\ell+1})\\
\hline
1&1&1&0\\
2&1&1&1\\
3&1&0&0\\
4&1&0&1
\end{tabular}
\end{center}

For the two cases 2 and 3, Eq. (\ref{separability}) holds for $\ell+1$ literals. 

Let us show that cases 1 and 4 are ruled out by the consistency equations. We have
\begin{equation}
\label{cases14} 
\Prob(i_2;\dots; i_\ell;i_{\ell+1})=\Prob(i_2)\dots \Prob(i_\ell)\cdot\Prob(i_{\ell+1}) =\Prob(i_1;i_2\dots; i_\ell;i_{\ell+1})+\Prob(-i_1;\dots; i_\ell;i_{\ell+1})
\end{equation}
where the first equality holds from the induction hypothesis for $\ell$ literals.

In case 4, Eq. (\ref{cases14}) reads
 $\quad 0 =1+\Prob(-i_1;i_2\dots; i_\ell;i_{\ell+1})$
 which is impossible.
 
In case 1,  still from Eq. (\ref{cases14}), $\Prob(-i_1;i_2\dots; i_\ell;i_{\ell+1})=\Prob(i_{\ell+1})=1$.
Similarly we obtain, $\Prob(i_1;-i_2;\dots; i_\ell;i_{\ell+1})=1$, etc. This contradicts the normalization, Eq. (\ref{normel}),
$$\sum_{2^{\ell+1}\mathrm{terms}}\Prob(\pm i_1;\pm i_2;\dots; \pm i_\ell;\pm i_{\ell+1})=1$$

 Therefore, only the two cases 2 and 3 are possible, so that Eq. (\ref{separability}) always holds for $\ell+1$ literals. 
$\Box$
\paragraph{}
This proves that a distribution separable over the set of working unknowns is consistently separable. Therefore, it defines a probability measure over the complete sigma-algebra $\mathcal{T}$.
\begin{proposition}
\label{propdeterministic}
Any deterministic solution on the set of working unknowns induces a deterministic probability measure on the sigma-algebra $\mathcal{T}$.
\end{proposition}
\emph{Proof.}
In general, all unknowns  of one literal, $\Prob(\pm i)$ with $i\in\llbracket 1, N\rrbracket$ are included in the set of working unknowns. As an exception, some unknowns $\Prob( i_0)$ may be absent, meaning that their values are indifferent. In this case we can assign whatever deterministic truth value to $\mathsf{X}_{i_0}$ for definiteness. Then, the truth value of any state $\omega$ in the sample set $\Omega$ can be computed by Eq. (\ref{separability}) and next the truth value of any event in the full sigma-algebra $\mathcal{T}$ as well.  By construction, this distribution is identical to the distribution already defined on the set of working unknowns.
$\Box$

\begin{remark}
\label{rmq2}
Technically, the deterministic solutions are the extreme points of the Bayesian simplex defined in Remark (\ref{rmq1}).
\end{remark} 

Proposition (\ref{propdeterministic}) has a corollary in the case of \emph{strict satisfiability} problems, where all variables are input variables that can be assigned independently to a truth value. Then the prior is a single Boolean function which determines the LP system, Eq. (\ref{lpeq}). Conversely, from  Proposition (\ref{propdeterministic}), the complete truth table of the Boolean function  is uniquely determined by the system Eq. (\ref{lpeq}), that is, the prior Boolean function is uniquely determined by Eq. (\ref{lpeq}). 
\begin{proposition}
\label{convdeterministic}
In a problem of strict satisfiability, the LP system Eq. (\ref{lpeq}) determines the truth table, that is the single Boolean function of the prior. 
\end{proposition}
\emph{Proof.} By definition of a problem of strict satisfiability, it is possible to assign any deterministic truth value to all unknowns of one literal. This determines the truth value to all working unknowns and to all states $\omega$ of the sample set $\Omega$ as well. Now, for each such assignment, if the $m$ equations of the linear system are satisfied, the truth value of the Boolean function in the prior is TRUE by Proposition (\ref{propdeterministic}). Otherwise, this truth  value is FALSE. Therefore, we obtain in principle, that is by \emph{force brute}, the complete truth table of the single Boolean function that acts as the prior. When the truth table is identically FALSE, this Boolean function is equal to $\varnothing$ by definition. $\Box$

\paragraph{}
In a problem of general satisfiability, the existence of deterministic solutions  must be checked by optimization with a convenient objective function. 
By contrast, thanks to Proposition (\ref{convdeterministic}), this difficulty is completely circumvented in problems of strict satisfiability, since the notion of \enquote{feasibility} of the LP system merges with that of \enquote{satisfiability} of the Bayesian prior Boolean function. This is specifically the case of the 3-SAT problem.

\subsection{Solving strict satisfiability problems}
\label{strictsat}

Consider a problem of strict satisfiability, specified by a single Boolean formula compelled to be valid, $\mathsf{f}=1$.

\begin{proposition}[Strict satisfiability]
When the prior is just a single Boolean function $\mathsf{f}$ compelled to be valid the problem accepts a deterministic solution if and only if the Bayesian LP system Eq. (\ref{lpeq})  is feasible.
\label{separable}
\end{proposition}
\emph{Proof}
Assume that the  prior depicts a single logical formula compelled to be valid, $\mathsf{f}=1$.
At least in principle, the truth table of $\mathsf{f}$ can be directly computed  from Proposition (\ref{convdeterministic}). Therefore, if $\mathsf{f}\not=\varnothing$, the system Eq. (\ref{lpeq}) accept a deterministic solution. Otherwise, $\mathsf{f}=\varnothing$ and there is no probability distribution compatible with $\Prob(\mathsf{f})=\Prob(\varnothing)=1$, so that the LP system is unfeasible. $\Box$
\paragraph{}
In other words, checking the existence of deterministic solutions does not require any optimization procedure.

\section{Polynomial time resolution of the 3-SAT problem}
\label{3sat}
We will now apply the present method to the resolution of the 3-SAT problem, as emblematic of the \textsf{NP} complexity class~\cite{cook, karp}.
\subsection{Description}
The 3-SAT problem is to determine whether a logical function, defined as a conjunction of $M$  disjunctions (or clauses)   with at most three literals per clause, is or not satisfiable. For instance, a particular clause $\mathsf{C}_r$ may be
\begin{equation}
\label{clauser}
\mathsf{C}_r\ident (\mathsf{X}_{i_r},\overline{\mathsf{X}}_{j_r},\mathsf{X}_{k_r})
\end{equation}
where $i_r,j_r,k_r\in\llbracket 1, N \rrbracket$ are distinct, $r\in\llbracket 1,M\rrbracket$.
This defines a Boolean function, $\mathsf{f}$, as the conjunction of the $M$ clauses.
With our terminology, this is a problem of strict satisfiability $\Prob(\mathsf{f})=1$.

\subsection{Specific equations}
\label{3satspec}
In order to account for the prior, each clause must be transcribed into a linear equation. For example, the clause Eq. (\ref{clauser}) compelled to be valid is transcribed as
$$\Prob({i_r},- j_r, k_r)=1$$
so that the conjunction of the $M$ clauses is translated into a system of $M$ linear equations.

It is convenient to use rather the negation of each clause, Eq. (\ref{negationl}), to obtain $M$ equations in terms of partial probability. For instance,  the validity of Eq. (\ref{clauser}) is retranscribed as
\begin{equation}
\label{clause}
\Prob(-i_r ; j_r; -k_r)=0,
\end{equation}
and we have by construction $M$ similar equations.

\emph{The 3-SAT problem is completely defined by $M$ \emph{specific equations} like Eq. (\ref{clause})}. 
This number of equations $M$ is always bounded by $M_\mathrm{max}=8\binom{N}{3}= \mathrm{O}(N^3)$ and even for non trivial problems, generally $M=$ O$(N)$.

\subsection{Working unknowns}
The working unknowns are derived from the $M$ clauses.  Due to possible duplication, we have a maximum of $M$ positive unknowns of 3 literals, bounded by $\binom{N}{3}$. As a result we have a maximum of $3M$ positive unknowns of 2 literals, bounded by $\binom{N}{2}$ and  a maximum of $3M$ positive unknowns of 1 literals, bounded by $\binom{N}{1}=N$.
 
Still due to possible duplication, the total number $n$ of working unknowns is $\le  26 M$. In any cases, this number  is polynomial in $N$ with  a maximum of $8\binom{N}{3}+4\binom{N}{2}+2\binom{N}{1}=$ O$(N^3)$ and generally $n=$ O$(N)$ for non trivial problems.

\subsection{Consistency equations}
The consistency equations are derived from the positive unknowns. The maximums are $12M$ bounded by $4\binom{N}{3}$ for 3 literals, $12M$ bounded by $2\binom{N}{2}$ for 2 literals and $3M$ bounded by $\binom{N}{1}$  for 1 literal, i.e., a total of $27M$ with possible duplication that is in fact $\le 27M$. 

\subsection{LP system}
The final LP system is defined on the $n$ working unknowns and $n$ is polynomial in $N$. 
Accounting for the $M$ specific equations and the consistency equations the total number of equations is $m\le 28 M$.
This number  is also polynomial in $N$ with  a maximum of $4\binom{N}{3}+2\binom{N}{2}+\binom{N}{1}=$ O$(N^3)$ and generally $m=$ O$(N)$ for non trivial problems.

\subsection{Satisfiability}
3-SAT is clearly a strict satisfiability problem. As a result, Proposition (\ref{separable}) holds.
\begin{proposition}[3-SAT satisfiability]
The 3-SAT problem accepts a deterministic solution if and only if the LP system Eq. (\ref{lpeq})  is feasible.
\label{3SATsat}
\end{proposition}

Since the dimension of the LP system is polynomial in the number of variables, its feasibility can be computed in polynomial time~\cite{khachiyan, karmakar}. Therefore,  the same conclusion applies to the satisfiability of the 3-SAT problem.

\begin{theorem*}
In the framework of Bayesian inference theory the 3-SAT problem can be computed in polynomial time in the number of variables.
\end{theorem*}
In other words, in the framework of Bayesian inference theory, 3-SAT is in \textsc{P}. 
Now, in the theory of algorithmic complexity~\cite{cook, levin}, the 3-SAT problem is \textsc{NP}-complete~\cite{cook, karp}, meaning that any \textsc{NP} language can be reduced to the 3-SAT problem in polynomial time. Eventually, we have the major achievement:
\begin{corollary*}
In the framework of Bayesian inference theory, \textsc{P} = \textsc{NP}.
\end{corollary*}

\begin{remark}
\label{rmq3}
This corollary seems to contradict a well-known property that finding integral solutions to a general LP system  is  \textsc{NP}-hard.
But, from Proposition (\ref{lemmedeterministic}), this property does not hold for Bayesian LP systems. 
\end{remark} 

\subsection{Search of the complete solution}
A particular solution can be computed by checking the feasibility of $N$ successive LP systems of decreasing dimension. The initial LP system of dimension $n$ checks the overall feasibility. If feasible,  we can determine a valid truth value to (say) $\mathsf{X}_N$. To this end, assign the truth value 1 to $\mathsf{X}_N$ and  check the feasibility of the new system of dimension less or equal to $n$. Then, if this new system is still feasible,  keep the assignment, otherwise,  change to the truth value 0. In any cases, this step provides both the truth value of $\mathsf{X}_N$ for a particular solution and a feasible system to determine the truth values of the $N-1$ other variables. The complete solution is finally obtained by iteration.

\section{Concluding remarks: A background for artificial neural networks}
The complexity of algorithms is due to the long combinatorial searches  which occur in the calculation, as a consequence of the discrete nature of the variables.
The Bayesian inference technique proves to be an effective loophole to circumvent this problem because continuous parameters, namely probabilities, are substituted for discrete variables,  allowing to replace effectively the intractable loops by a smooth optimization process. 
Remarkably, even this optimization process is not necessary for problems of strict satisfiability. As a result, there is no difference between \textsf{P} and \textsf{NP} problems and any decidable problem of class \textsf{NP} can be solved in polynomial time. The point is to use probability instead of discrete symbols.

But there are at least two other standard techniques that use probability.

 The first is quantum computing where probabilities are defined in a convenient complex Hilbert space. This also provides a dramatic speedup of the calculation.
 However, the few known quantum programs are quite sophisticated randomized algorithms and, above all, require a  very highly problematic  setup~\cite{ibm, google,dyakonov}.  
By contrast, the present framework uses a single inference technique to  solve every problem and requires no particular setup. 

The second is artificial intelligence. This technique is in fact closely connected to Bayesian inference, namely, estimate the likelihood of some conclusion given a set of prior data.
 Indeed,  natural neurons proceed by Bayesian inference (see. e.g. \cite{ma}).
 Furthermore, as pointed out by E. T. Jaynes \cite{jaynes1}, Bayesian inference appears as the relevant background for implementing artificial neural networks.
 Now, the present paper proposes a new way to address this approach, highlighting for example the natural emergence of logical layers. We use linear programming for completing the computation but above all to prove that the complexity classes vanish in this context. 
In reality, linear programming is not necessarily the more powerful means of resolution.
Bypassing linear programming,  artificial neural networks should indeed be the optimum method to solving any \textsf{NP} problem in polynomial time.

\phantomsection
\addcontentsline{toc}{section}{References}
\bibliography{biblio}

\end{document}